\documentclass{article}
\usepackage{graphicx} 
\usepackage{todonotes}
\usepackage{hyperref}
\usepackage{xcolor}

\title{Large language models in 6G security: challenges and opportunities}

\author{Tri Nguyen$^1$, Huong Nguyen$^1$, Ahmad Ijaz$^2$,\\ Saeid Sheikhi$^1$, Athanasios V. Vasilakos$^3$, Panos Kostakos$^1$}

\date{$^1$ University of Oulu, Finland \\
$^2$ VTT, Finland \\
$^3$ University of Agder, Norway}

\begin{document}

\maketitle

\begin{abstract}
The rapid integration of Generative AI (GenAI) and Large Language Models (LLMs) in sectors such as education and healthcare have marked a significant advancement in technology. However, this growth has also led to a largely unexplored aspect: their security vulnerabilities. As the ecosystem that includes both offline and online models, various tools, browser plugins, and third-party applications continues to expand, it significantly widens the attack surface, thereby escalating the potential for security breaches. These expansions in the 6G and beyond landscape provide new avenues for adversaries to manipulate LLMs for malicious purposes.

We focus on the security aspects of LLMs from the viewpoint of potential adversaries. We aim to dissect their objectives and methodologies, providing an in-depth analysis of known security weaknesses. This will include the development of a comprehensive threat taxonomy, categorizing various adversary behaviors. Also, our research will concentrate on how LLMs can be integrated into cybersecurity efforts by defense teams, also known as blue teams. We will explore the potential synergy between LLMs and blockchain technology, and how this combination could lead to the development of next-generation, fully autonomous security solutions. This approach aims to establish a unified cybersecurity strategy across the entire computing continuum, enhancing overall digital security infrastructure.

Our comprehensive analysis, drawing from academic research, proof-of-concept studies, and renowned cybersecurity resources like OWASP, aims to equip LLM stakeholders with a detailed, actionable road-map. This guide focuses on enhancing defense strategies informed by threats to LLM applications. Furthermore, the development of a threat taxonomy, specifically for Generative AI and LLMs, will significantly enhance the robustness of novel frameworks like the AI Interconnect. By categorizing potential adversarial behaviors, this taxonomy empowers the framework to proactively address security vulnerabilities, thereby strengthening the security and resilience of the 6G network ecosystem.
\end{abstract}
\section{Introduction}



The swift adoption of Generative AI (GenAI) and Large Language Models (LLMs) across diverse sectors, including education and healthcare, signifies a monumental leap forward in technological innovation. These advancements offer unprecedented opportunities for enhancing learning experiences, streamlining information processing, and facilitating more effective healthcare solutions. Yet, alongside these benefits, the rapid proliferation of GenAI and LLMs has unveiled a critical, yet often overlooked, dimension: the emergence of security vulnerabilities.
As the ecosystem encompassing both offline and online models continues to grow, incorporating a myriad of tools, browser extensions, and third-party applications, the potential for security risks escalates correspondingly. This expansion not only broadens the attack surface but also introduces complex challenges in safeguarding these technologies against exploitation. In the era of 6G and beyond, where connectivity and computational power are greatly enhanced, the avenues for adversaries to infiltrate and manipulate LLMs for nefarious purposes have multiplied.
This evolving landscape necessitates a concerted effort to address these security concerns, ensuring the safe and ethical utilization of GenAI and LLMs. As we navigate these advancements, it becomes imperative to develop robust security measures and protocols that can shield these technologies from potential threats, safeguarding the integrity of the innovations that stand to revolutionize sectors as vital also as the use to develop 6G.

In this section, our examination zeroes in on the security dimensions of LLMs, through the lens of potential adversaries. We delve into their aims and tactics, aiming to provide a thorough analysis of recognized security vulnerabilities associated with LLMs. Our exploration will unfold a detailed threat taxonomy that classifies various adversarial behaviors, offering insights into the array of security challenges.
Furthermore, our investigation extends to the strategic incorporation of LLMs into cybersecurity measures undertaken by defensive teams, commonly referred to as blue teams. This integration is pivotal for bolstering defense mechanisms against sophisticated cyber threats.
Building upon this foundation, we introduce and consider the concept of LLMSecOps inspired from Security Operations (SecOps) within practical scenarios, with a particular emphasis on its relevance and application in the burgeoning 6G landscape. An integral part of our discourse also revolves around the innovative convergence of LLMs with blockchain technology. We posit that this fusion holds the promise of pioneering next-generation, autonomously operating security solutions. 
Our objective is to craft a comprehensive cybersecurity strategy that spans the entire computing spectrum. By doing so, we aim to significantly reinforce the digital security infrastructure, ensuring a robust defense against emerging and evolving cyber threats.

Our comprehensive analysis, drawing from academic research, proof-of-concept studies, and renowned cybersecurity resources like the Open Web Application Security Project (OWASP), aims to equip LLM stakeholders with a detailed, actionable road map. This guide focuses on enhancing defense strategies informed by threats to LLM applications. Furthermore, the development of a threat taxonomy, specifically for GenAI and LLMs, will significantly enhance the robustness of novel frameworks like synergizing LLMs or autonomous LLM agent swarms. By categorizing potential adversarial behaviors, this taxonomy empowers the framework to proactively address security vulnerabilities, thereby strengthening the security and resilience of the 6G network ecosystem.


\section{Attacks - Red teaming}



In this section, we will conduct a thorough exploration of the current vulnerabilities in the field and develop a comprehensive taxonomy, differentiating between the types, objectives, and strategies. This taxonomy will be instrumental in informing and guiding the application of LLMs in the 6G computing continuum. 

Recently, OWASP has convened a multidisciplinary team of experts from a broad spectrum of disciplines, such as security, Artificial Intelligence (AI), software development, and industry leadership~\cite{OWASP10LLM}. This coalition's objective is to systematically identify and underscore the critical security and safety challenges that both developers and security professionals need to be aware of when they are integrating LLMs into application development. Below is the preliminary compilation of critical vulnerability categories related to AI applications developed using LLMs: 
1) Prompt Injection, 2) Insecure Output Handling, 3) Threats of Training Data Poisoning, 4) Model Denial of Service Attacks, 5) Supply Chain Concerns, 6) Disclosure of Sensitive Information, 7) Design Flaws in Insecure Plugins, 8) Excessive Agency in Models, 9) Overreliance on AI Models, 10) Model Theft.

Furthermore, numerous review studies have also aimed to explore the limitations, challenges, potential risks, and opportunities presented by GenAI in the realms of cybersecurity and privacy \cite{gupta2023chatgpt,sultana2023towards}. According to Yao et al.~\cite{yao_2023_survey}, these vulnerabilities can broadly be categorized into two main types: AI-inherent vulnerabilities and non-AI inherent vulnerabilities.

\subsection{AI-related vulnerabilities} 
This type of vulnerability refers to the susceptibility or weakness inherent to LLMs' design, architecture, or behavior. These vulnerabilities arise due to the complex nature of LLMs and the challenges associated with training, deploying, and managing them in real-world applications. Furthermore, recent findings indicate that deceptive (i.e., backdoor) behavior, especially in the larger models and those trained with chain-of-thought reasoning, can be made persistent, evading elimination through conventional safety alignment techniques, such as supervised fine-tuning, reinforcement learning, and adversarial training \cite{hubinger2024sleeper}. 

    \begin{enumerate}
        \item \textbf{Adversarial attacks:} This attack aims to deceive the model and manipulate the input data thereby affecting model performance. Two primary strategies of this type include data poisoning~\cite{wan2022you, schuster2021you, shan2023prompt} and backdoor attacks~\cite{li2023chatgpt, yao2023poisonprompt, yang2023comprehensive}. The key difference between these two is that backdoor attacks specifically embed hidden triggers into the model to manipulate specific behaviors while poisoning attacks directly inject malicious examples into the dataset to deceive the model and hinder the training process.
        
        \item \textbf{Inference attacks:} The attack aims to deduce sensitive information or insights about the model and its training data by making specific queries or observations, waiting for unintended information leakage from the responses to exploit. Several subset attacks can be viewed, such as attribute inference attacks~\cite{kandpal2023user, song2020information, staab2023beyond} and membership inferences~\cite{shokri2017membership, carlini2022membership, choquette2021label,duan2023diffusion}. In both types, the term ``inference" refers to the process of deriving information from the model's outputs or behavior, albeit with distinct kinds of information being extracted in each. Given their names, the attribute inference attack allows attackers to infer attributes such as demographics, preferences, or characteristics present in the data while membership inference attacks aim to determine whether a specific data record was a part of the training dataset for an LLM. The attackers seek to identify if a specific piece of data (e.g. an individual's record) was used during the model's training process. This attack is concerning because it can reveal sensitive information about individuals, breach privacy, and compromise data confidentiality. If successful, attackers could infer membership in sensitive groups or datasets, leading to privacy violations, identity theft, or unauthorized access to confidential information. 
        
        \item \textbf{Extraction attacks:} This type of attack, while similar to inference attacks regarding disclosing sensitive information or insight into learning models, differs in their methods and targets. Specifically, extraction attacks aim to obtain specific resources or confidential information directly, such as model gradients or training data, whereas inference attacks typically involve observing the model's responses or behavior to deduce information from the dataset. Primary approaches in this type include model stealing~\cite{juuti2019prada, kariyappa2021maze},  gradient leakage~\cite{balunovic2022lamp}, and training data extraction~\cite{carlini2021extracting}, and \cite{Truong_2021_CVPR} points out the possibility to replicate the model without accessing the original model data.

        \item \textbf{Bias and unfair exploitation:} Bias and unfairness in LLMs arise from systematic errors in language generation, often stemming from biases present in training data or introduced during model fine-tuning~\cite{talat2022you}. If a training dataset disproportionately represents certain demographics or perspectives, the model may inadvertently learn and perpetuate biases in its generated text. This biased language can result in misinformation~\cite{su2023fake, urman2023silence}, reinforcing stereotypes, societal inequalities, and discriminating against certain groups (e.g. by gender and minority groups~\cite{kotek2023gender, urchs2023prevalent}). 

        \item \textbf{Instruction tunning attack:} The attack manipulates LLMs to execute unintended actions or surpass system limitations. Notable exploits include Denial of Service - DoS~\cite{namer2024automatically}, which targets system availability by overloading it with excessive requests; indirect prompt injection~\cite{abdelnabi2023not, liu2023prompt, liu2023promptattndef}, crafting inputs to induce unintended or harmful responses; and jailbreaking~\cite{taveekitworachai2023breaking, deng2023jailbreaker}, aiming to bypass constraints and enable restricted functionalities. Particularly, the prompt injection expands to various attacks, including virtualizing prompt injection~\cite{yan2023backdooring}, unveiling guide prompt~\cite{zhang2023prompts}, and integrating application~\cite{liu2023prompt}.

        \item \textbf{Zero-day attacks:} Prior research has demonstrated that the inability to eliminate backdoors from models using alignment methods may create vulnerabilities similar to ``sleeper agents". In this scenario, an attack, such as data theft, could be covertly embedded within the model's weights. This type of attack could be activated at will by specific trigger phrases, which could range from a sequence of strings to UTF-8 characters or even Base64 encoded messages \cite{hubinger2024sleeper}.

    \end{enumerate}

\subsection{Non-AI-related vulnerability} 
In comparison, attacks in this category are more related to the system-level risks and other associated plugins, not directly to the function of the model (e.g. interfering in training or response). 

    \begin{enumerate}
        \item \textbf{Remote Code Execution (RCE):} RCE involves remotely executing arbitrary code on app servers by exploiting vulnerabilities in software or services via prompt injection. Attackers may execute various types of malicious code or commands, including (1) creating backdoors for persistent access and control without the user's knowledge, (2) exfiltrating data and sending it back to attackers, (3) escalating privileges for more extensive control and access to restricted areas, or (4) disrupting operations by deleting critical files and corrupting data. While the specific code or commands in an RCE attack may vary in approaches, they ultimately serve the same goals of data theft, system control, and disruption~\cite{liu2023demystifying}.

        \item \textbf{Side channel:} The goal of side-channel attacks in LLMs is often to gather the information via practical deployments that can be used to compromise the system or model further, whether for unauthorized access, data exfiltration or to replicate the model's capabilities. These attacks do not directly exploit vulnerabilities in the LLM itself but instead analyze patterns, physical phenomena (e.g. timing, power consumption), or other observable characteristics to gain insights that should otherwise not be accessible~\cite{debenedetti2023privacy}. Understanding this system information, attackers can analyze power consumption and other side-channel information to infer system workloads, identifying when a system processes complex tasks or is vulnerable due to operational strain~\cite{nair2023hardened}. This insight also aids in timing attacks or reverse engineering by revealing model complexities.

        \item \textbf{Insecure plugin:} In this attack, attackers might target plugins supported in the LLM ecosystem but have vulnerabilities due to poor security practices, lack of updates, or inherent design flaws~\cite{drori2024human}. Beyond exploiting the existing plugins,  attackers can extend their tactics by deploying their own, aiming to manipulate the LLMs' behaviors, extract sensitive data, steal chat histories, access personal information, or execute code on users’ machines, often leveraging OAuth vulnerabilities in plugins~\cite{OWASP10LLM}.
    \end{enumerate}

\section{Defense - Blue teaming}
In the last few years, the exploration of LLMs for cybersecurity operations has significantly advanced. Yao et al. \cite{yao_2023_survey} conducted a detailed examination and analysis of 279 research papers from 2021 to 2023, investigating the relationship between LLMs and security and privacy concerns before pointing out strategies for LLM training safety. 
Moreover, to enhance our understanding of the versatility of LLMs across different cybersecurity operations, this section provides a review of existing research results applicable to cyber defense (i.e., blue) teams. 

\subsection{Strategies for LLM training safety}
    Due to the border security aspects of non-AI inherent vulnerabilities, \cite{yao_2023_survey}'s discussion focuses on strategies to improve LLM training safety. Below is the set of strategies to mitigate LLM vulnerabilities regarding AI inherent.
    \begin{itemize}
        \item \textbf{LLM training:} 
        The development of LLMs encompasses intricate decisions regarding model architectures, the selection and preparation of training data, and the adoption of specific optimization techniques. Each of these components plays a crucial role in ensuring the security and privacy of LLMs throughout their lifecycle.
            \begin{itemize}
                \item \textbf{Model architectures and privacy preservation:} The management of storage and organization within LLM architectures is paramount for maintaining data privacy. Recent studies have shown the effectiveness of incorporating differential privacy techniques during the training phase to safeguard user data~\cite{li2022large}. Additionally, enhancing LLMs' resilience against adversarial attacks has been a focus of ongoing research~\cite{zhu2023promptbench}.
                \item \textbf{Knowledge integration for enhanced trust:} Incorporating external knowledge sources, such as knowledge graphs~\cite{zafar2023building}, into LLMs can significantly improve their trustworthiness and cognitive robustness~\cite{romero2023synergistic}. These enhancements not only contribute to the models' understanding of complex concepts but also bolster their defenses against misleading information.
                \item \textbf{Corpora cleaning for bias reduction: } The quality of the training corpora is fundamental to preventing bias and ensuring high-quality data input. Rigorous corpora cleaning processes are essential for eliminating biases and enhancing the contextuality and accuracy of the training data~\cite{Ganesh_2023_on,ousidhoum-etal-2021-probing}.
                \item \textbf{Optimization techniques for secure learning:} Optimization strategies influence how LLMs interpret and learn from data, directly impacting their security. Adversarial training methods have been developed to train LLMs to withstand malicious inputs~\cite{ivgi2021achieving,yoo2021improving}. Furthermore, aligning LLMs' objectives with safety principles through human feedback has emerged as a promising approach to mitigate unintended harmful behaviors~\cite{NEURIPS2022_b1efde53,yuan2023rrhf}.
            \end{itemize}
        
        \item \textbf{LLM inference:} Deploying LLMs within systems that interact with users in real time necessitates a comprehensive security strategy. This strategy should encompass three critical phases: prompt pre-processing, abnormal detection, and response post-processing. By meticulously implementing safeguards at each phase, we can significantly enhance the security of LLM interactions, potentially unlocking new possibilities for LLM connectivity and distributed applications.
        \begin{itemize}
            \item \textbf{Prompt pre-processing: } The initial phase focuses on mitigating risks from potentially malicious user inputs, commonly associated with jailbreak attacks. Strategies include: (1) Instruction Manipulation Prevention: Implementing checks to identify and neutralize attempts to alter instructions in a way that could compromise the system~\cite{kirchenbauer2023reliability}; (2) Defensive Demonstrations: Utilizing examples of secure and compliant interactions to guide the LLM away from fulfilling harmful requests~\cite{wei2023jailbreak}; (3) Purification of Inputs: Applying techniques to cleanse input data of any elements that could lead to undesirable outputs~\cite{li2023text}.
            \item \textbf{Malicious detection: } This phase involves a thorough analysis of the LLM's outputs based on the given inputs to identify any prompt injection threats or backdoored instructions. Recent advancements include: (1) Backdoored Instruction Detection: Techniques proposed by Sun et al.~\cite{Sun_Li_Meng_Ao_Lyu_Li_Zhang_2023} focus on identifying hidden malicious commands embedded within seemingly benign inputs; (2) Abnormal and Poisoned Instruction Detection: \cite{xi2024defending} introduce methods to detect and mitigate the impact of abnormal or poisoned instructions, ensuring the integrity of the LLM's outputs.
            \item \textbf{Response post-processing:} Before presenting the generated responses to users, a final verification step is crucial, particularly, studies ~\cite{xiong2023llms,phute2023llm}, pointing assessing harmfulness and confidence, suggest mechanisms for evaluating the potential harm and reliability of LLM responses, ensuring that outputs are both safe and contextually appropriate.
            
        \end{itemize}

    \end{itemize}

\subsection{Taxonomy and LLMSecOps applications}




Specifically, Sultana et al. \cite{sultana2023towards} systematically assessed the role of LLMs in cyber operations, exploring their utility in key cyber defense tasks. By analyzing literature on network defense, including Cyber Threat Intelligence (CTI) analysis, log management, anomaly detection, and incident response, they developed a four-tier taxonomy. This taxonomy organizes the extensive capabilities of LLMs into distinct cyber operations categories, closely aligned with the widely adopted version 1.1 of the National Institute of Standards and Technology (NIST) Cybersecurity Framework \cite{nist}.
\begin{itemize}
    \item \textbf{Identify} focuses on LLMs for identifying and classifying threats from open-source CTI, enhancing early threat intelligence efforts.
    \item \textbf{Protect} uses LLMs for vulnerability scans, security assessments, and automating defense strategies, bolstering network security.
    \item \textbf{Detect} applies LLMs to detect vulnerabilities, extract malware, and classify attacks, highlighting their role in early threat identification.
    \item \textbf{Respond} leverages LLMs in incident response and recovery, aiding in analysis and strategic planning post-incident.
\end{itemize}

Furthermore, the section presents an in-depth analysis of diverse systems and tools designed to leverage LLMs with the aim of enhancing cybersecurity operations, across the Identify, Protect, Detect, and Respond phases of the NIST framework. This exploration is focused on understanding how these technological advancements are integrated and their significant contribution to the cybersecurity domain. We specifically focus on demonstrating the capabilities of LLMs within the SecOps framework, emphasizing their role in the advancing 6G edge-cloud continuum. The section highlights the creative applications of LLMs in strengthening security infrastructures and enhancing response strategies.

\textbf{PentestGPT}: Deng et al. \cite{deng2023pentestgpt} developed PentestGPT, an automated penetration testing tool powered by LLMs such as GPT-3.5 and GPT-4, aimed at enhancing cybersecurity tasks. This tool \cite{GreyDGL2023} is specifically engineered to optimize the penetration testing process, offering an interactive platform that supports penetration testers by delivering insights on their progress and detailed attack vectors. To evaluate the tool, the authors established a comprehensive benchmark incorporating test environments from HackTheBox and VulnHub, which are prominent platforms for penetration testing Capture-The-Flag (CTF) challenges. This benchmark was designed to assess the tool's performance across 26 types of sub-tasks critical to penetration testing, including Port Scanning, Web Enumeration, Code Analysis, Shell Construction, Directory Exploitation, General Privilege Escalation, Flag Capture, Password/Hash Cracking, and Network Exploitation, incorporating active human involvement in the loop for enhanced efficacy. Leveraging prompt tuning, PentestGPT demonstrated significant performance improvements over vanilla models, achieving a 228.6\%  and 58.6\% increase in the completion of these sub-tasks when compared to the performance levels when directly using GPT-3.5 and GPT-4, respectively.

\textbf{PAC-GPT}: Kholgh and Kostakos \cite{kholgh2023pac} introduced a novel framework leveraging OpenAI's GPT-3 to create synthetic network traffic, aiming to fill the gap in the availability of realistic cybersecurity datasets. The primary goal of this research was to generate high-quality synthetic data that can be used for training and evaluating cybersecurity systems, such as intrusion detection systems. The methodology centered on developing a Flow Generator and Packet Generator, which utilized GPT-3's advanced capabilities with minimal fine-tuning, and a user-friendly command-line interface tool \cite{pacgpt} to facilitate various cybersecurity tasks. PAC-GPT  showed high levels of accuracy in generating synthetic network traffic that closely resembles real-world normal and malicious network activities. The success rates for packet generation in both normal and malicious scenarios indicate the potential utility of LLM to operate directly within network environments. As GenAI increasingly plays a pivotal role in facilitating the development of synthetic attacks \cite{electronics13020322}, PAC-GPT has shown that the generation/injection of pcap data into the network is becoming an important facet as well. 

\textbf{TSTEM:} Balasubramanian et al. \cite{tstem} developed the TSTEM platform, which is designed for real-time collection, processing, and dissemination of CTI from diverse online sources. TSTEM leverages a microservice architecture, incorporating tools like Tweepy, Scrapy, Terraform, ELK, Kafka, and MLOps for efficient, autonomous operation. The platform introduces an innovative, containerized approach to manage compute-intensive data pipelines within cloud computing environments, enhancing the process of identifying and indexing Indicators of Compromise (IOC). It utilizes LLMs, specifically employing cutting-edge models such as Bidirectional Encoder Representations from Transformers (BERT)~\cite{devlin2019bert}, Longformer~\cite{beltagy2020longformer}, and BERT-NER\footnote{huggingface.co/dslim/bert-base-NER} for the identification and extraction of CTI. Experimentation demonstrates exceptional performance, with accuracy exceeding 98\% in classification and extraction within a minute, illustrating its effectiveness in precise, low false-positive IOC identification across multiple stages. This platform not only addresses technical bottlenecks in CTI collection but also provides an open-source solution for deploying streaming CTI data pipelines, significantly contributing to cybersecurity defenses. The study builds upon a growing body of work that explores the application of LLMs to enhance the processing of cyber threat intelligence gathered from open sources \cite{alves2022leveraging,sultana2023towards}.  This research emphasizes the potential of LLMs to refine how cybersecurity threats are detected, analyzed, and mitigated, aiming to significantly reduce response times and improve the overall security posture. What distinguishes TSTEM from other initiatives is its practical application of LLMs within a microservice architecture, enabling seamless integration into the edge-cloud continuum.

\textbf{GPT-2C:} Setianto et al. \cite{setianto2021gpt}  developed a prototype leveraging a transformer model fine-tuned for analyzing honeypot logs, aiming to improve the capabilities of Intrusion Detection Systems (IDS).  By fine-tuning GPT-2 with a dataset of illegal Unix commands from the Cowrie SSH honeypot, the model achieved an 89\% inference accuracy. This approach offers a significant improvement in parsing unstructured log data, aiming to reduce the response time between threat detection and system patching. GPT-2C exemplifies the integration of AI in cybersecurity, providing a novel method for real-time log analysis with potential applications in advanced defense strategies.

\textbf{LogBERT:} \cite{guo2021logbert} is a self-supervised framework for detecting anomalies in system logs using BERT. The methodology for training LobBERT includes two novel training tasks: masked log message prediction and volume of hypersphere minimization, to learn normal log sequence patterns. LogBERT outperforms state-of-the-art approaches in anomaly detection across three log datasets, showcasing its effectiveness in capturing normal patterns and detecting deviations. 

\textbf{LogPPT:} \cite{le2023log} is a log parser leveraging prompt-based few-shot learning for converting raw log messages into structured data, addressing the limitations of existing log parsers by capturing semantic information of log messages with minimal labeled data. Utilizing RoBERTa\footnote{huggingface.co/docs/transformers/model\_doc/roberta}, a pre-trained language model, for virtual label token prediction, LogPPT achieves high accuracy and robustness across diverse log datasets without extensive preprocessing or domain-specific knowledge. Experimental results on 16 public log datasets demonstrate LogPPT's superior performance in parsing accuracy and efficiency compared to state-of-the-art methods.

\textbf{LogBot}: Balasubramanian et al \cite{balasubramanian2023transformer} proposed a method for enhancing chatbot performance in cybersecurity through anomaly detection, utilizing GPT-3 models combined with rule-based logic. The anomaly-aware agents developed for the study showcases the effectiveness of this approach in real-world scenarios, demonstrating over 99\% accuracy in anomaly detection within log files and outperforming other LLMs like BERT, DistilBERT~\cite{sanh2020distilbert}, and ALBERT~\cite{lan2020albert}. It highlights the critical need for integrating conversational agents with anomaly detection capabilities for analyzing complex log data, involving an iterative fine-tuning process of GPT-3 models to adapt them for anomaly detection. The paper also discusses the architecture and methodology of the proposed anomaly detection system, emphasizing its potential to set a new standard in conversational, anomaly-aware agents in cybersecurity.

\textbf{Cyber Sentinel:} \cite{kaheh2023cyber} is a cybersecurity dialogue system that utilizes GPT-4 to articulate potential cyber threats and implement both proactive and reactive security measures upon user command. The system merges LLM conversational agents with an IOC signature database hosted on Elasticsearch, and incorporates a Wazuh Security Information and Event Management (SIEM) module for a comprehensive security solution. The approach is grounded in the integration of GPT-4 models through strategic prompt engineering to facilitate task-specific conversations. The study shows the system's ability to use LLMs for better cybersecurity measures, marking a key advancement in Explainable and Actionable AI for SecOps. 

\textbf{HuntGPT:} \cite{ali2023huntgpt} is a specialized intrusion detection dashboard that leverages a Random Forest classifier trained on the KDD99 dataset, incorporating explainable AI (XAI) frameworks like SHAP and Lime to improve model transparency and user understanding. By integrating GPT-3.5 Turbo as a conversational agent, HuntGPT delivers easily understandable threat detections, aiming to enhance user experience and decision-making in cybersecurity operations. The paper evaluates the system's architecture, its technical accuracy using Certified Information Security Manager (CISM) Practice Exams, and the quality of response readability. Results suggest that conversational agents powered by LLM technology, when combined with XAI, can provide a robust mechanism for generating actionable AI solutions in the realm of intrusion detection systems.

\section{LLMSecOps in the 6G era}

The integration of AI into communication networks, notably within the 6G era, signifies a transformative shift towards ``AI Interconnect'' and ``AI-native Telecom'' paradigms or autonomous LLM agent swarms. This evolution, as highlighted in recent studies~\cite{britto2023telecom, tarkoma2023ai}, introduces a dual spectrum of vulnerabilities: those inherent to AI and those not specific to AI technologies~\cite{yao_2023_survey}. In addressing these challenges, the insights provided by OWASP~\cite{OWASP10LLM} and the governance surrounding LLMs emerge as pivotal considerations for advancing secure communication networks. The incorporation of LLMs marks a significant advancement in the telecommunication sector's vision. Nevertheless, ensuring the security and trustworthiness of LLM usage necessitates enhanced verification measures.


As research on the 6G computing continuum is still in its early phases, there is an expectation that the resulting reference architecture will integrate Key Enabling Technologies (KETs) like Network Function Virtualization (NFV), edge intelligence, Software-Defined Networking (SDN) across 5G Core, Cloud, and Edge networks. Furthermore, the system's architecture will be aligned with international standards, ensuring not only its full autonomy but also seamless interoperability with existing legacy systems \cite{taleb20226g}. In this scenario, LLMs will offer support for intelligent decision-making, evolving SecOps into fully autonomous cognitive LLMSecOps services seamlessly integrated within network functions.

\subsection{Intent-based networking}
\begin{figure}[h]
\centering
\includegraphics[width=1\textwidth]{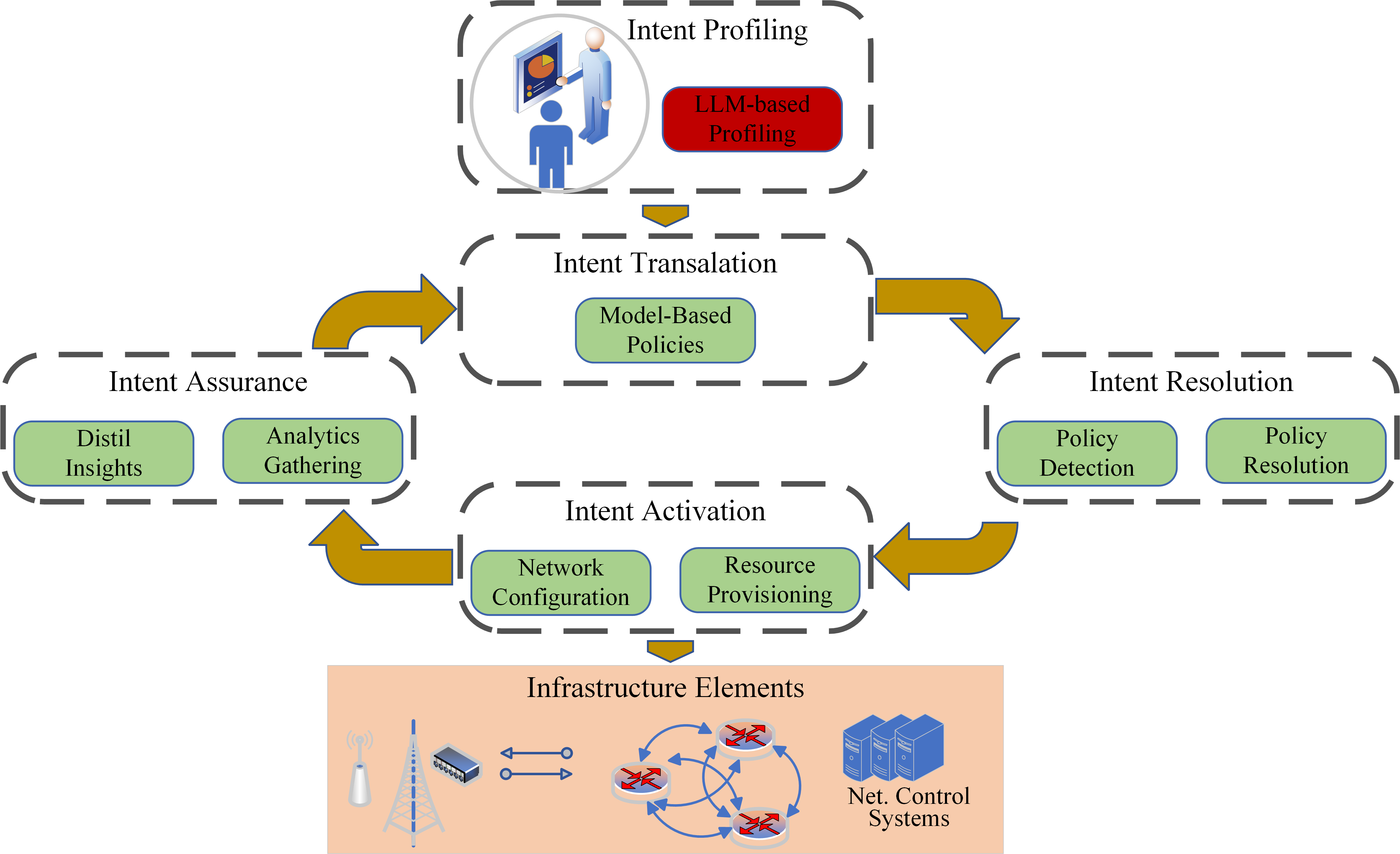}
\caption{Process and components of IBN.}
\label{fig:ibn}
\end{figure}

Intent-based networking (IBN)~\cite{9925251} is an emerging paradigm that aims to automate network configurations using AI. The basic idea of IBN is to enable network administrators to manage complex networks with business intents. Business intents can be high-level business objectives of organizations. For example, an organization can prioritize one type of traffic, such as video traffic, over other types of traffic, such as text. IBN requires developments in AI, such as LLMs, to effectively convert such intents into configurations. 6G will utilize LLMs to enable run-time network configurations through high-level intents to simplify human-network interactions and smoothen the deployment of new services. However, IBNs can have several security challenges~\cite{ahmad2023security}, that can be described along with the process or workflow of IBNs. The Internet Engineering Task Force (IETF) describes intents to be abstract, declarative, and vendor agnostic set of rules that can be deployed through several steps of a properly defined process~\cite{clemm2020intent}. A pictorial presentation of the process and flow of IBN is presented in Fig.~\ref{fig:ibn}.

The first component or process in IBNs is intent profiling where a network administrator expresses intent to an IBN system. This process must be user-friendly and the system must facilitate the user for meaningful intents. The second step is intent translation or compilation, where intents are transformed into low-level network configurations. LLMs can play a major role in these steps, to effectively transform service requests into network configurations. A major security risk here is that if the LLMs are compromised, the whole network can be compromised, for instance through malignant configurations. Sensitive traffic in this case can be diverted to malicious nodes for compromising privacy and security. However, if the next step, intent resolution is properly carried out, miss-configurations can be recognized. Since miss-configurations are responsible for a majority of network security challenges arising from human-network interactions~\cite{8712553}, LLMs-based configurations can minimize such vulnerabilities, given correct intent profiling carried out through LLMs. The next steps in IBN, intent activation which provides the necessary services intended by intents, and intent assurance which monitors the intents throughout its life cycle to ensure that the network behaves as intended, depend on the initial steps. Therefore, the security of LLMs-based intent profiling and translation is extremely important for IBN, and thus 6G. 

\subsection{Network Data Analytics Function}
The 3GPP Network Data Analytics Function (NWDAF) \cite{etsi_5g_2019}, envisioned as a key component of the 6G computing continuum's service base \cite{tarkoma2023ai}, is the analytics hub that establishes a foundation for AI/ML-driven data analytics operations and services \cite{garcia2023network}. This hub aims to ensure their complete integration and interoperability across the network. NWDAF is designed to aggregate and analyze data related to network efficiency, User Equipment (UE) behavior, service usage patterns, network Operations, Administration, and Maintenance (OAM) spanning the computing continuum, including WiFi, WAN, 5G Core, Cloud, and Edge networks \cite{etsi_5g_2021}. Specifically, this function is designed to perform analytical inferences, provide services for training machine learning models, and make data accessible via the \textit{Nnwdaf\_AnalyticsInfo} service.  This enables native downstream LLMSecOps services to acquire targeted analytics from NWDAF \cite{garcia2023network}.

\subsection{Zero-Touch Network 5G/6G Security}
The Zero-touch Network and Service Management (ZSM) is focused on revolutionizing network management towards a fully automated, flexible, and efficient approach. This extends across all mobile network generations, emphasizing the necessity for networks to autonomously self-configure, self-monitor, self-heal, and self-optimize without human intervention \cite{de2021end}. This automation is crucial as networks evolve, especially with the complexity introduced by 5G and future 6G technologies, supporting a vast number of connected devices and services with diverse network requirements \cite{de2021survey,sheikhi2023ddos}. ZSM contains advanced technologies like AI and ML for intelligent decision-making, as well as SDN and NFV for simplifying network management, indicating a shift towards networks that can independently adapt to and defend against cyber threats, ensuring robust and resilient future telecommunications infrastructure

The integration of Zero-touch Network Management (ZTM) into the ZSM framework is essential to improve network management's security and efficiency in 5G and 6G networks. The ZSM initiative aims for full end-to-end network automation, and ZTM plays a key role in achieving this goal \cite{gallego2022machine}. The framework can enable end-to-end network slicing and AI-based security mechanisms, ensuring a secure infrastructure that caters to diverse service requirements. Furthermore, the adoption of AI and ML optimizes network operations and introduces new security paradigms, addressing challenges posed by SDN, NFV, MEC, and network slicing. These advancements are crucial for the creation of secure, autonomous systems capable of proactively identifying and mitigating threats, thus safeguarding the integrity and reliability of future mobile networks \cite{coronado2022zero}.

The anticipated complexity of future 6G networks will escalate to levels where conventional analytical and numerical simulation methods will become impractical \cite{coronado2022zero}. This necessitates the ZSM framework reference architecture to embed data protection mechanisms for use, transit, and storage. This ensures an elevated security standard across management functions, services, and infrastructure resources while safeguarding data security and integrity \cite{liyanage2022survey}. Therefore, integrating Zero-Touch with LLMs can enhance the security of 6G networks. This integration could help create a future where networks can autonomously defend against cyber threats through intelligent, adaptive, and automated mechanisms \cite{mahmoodi2023autonomous}. These systems would continuously learn from new data, adapt to emerging threats, and implement security measures without manual configuration or intervention \cite{sheikhi2023cyber}. As a result, a robust and resilient 6G infrastructure can be ensured.

    \section{Autonomous LLM Agent Swarms}

    This section delves into the current intersection of defense applications employing LLMs in a decentralized setting. It leverages these cutting-edge technologies as a foundation to propose a forward-looking perspective on the future landscape of distributed LLM or LLM multi-agent systems. This investigation is pivotal in supporting the AI Interconnect's commitment to embedding security and trust principles from the outset, thereby crafting a more robust and trustworthy framework for future technological deployments.

    \subsection{The transition to distributed LLMs} 
    The prevailing model for LLMs is predominantly centralized, managed by singular organizations. This centralization introduces critical challenges, including discrepancies in model design and the utilization of potentially biased or sub-optimal training data. To address these issues, a decentralized architecture for LLMs is proposed, leveraging a network of LLMs to validate responses and offer a diversity of perspectives. This decentralization not only mitigates the concerns associated with centralized models, such as privacy risks, usage restrictions, and dataset biases, but also fosters greater openness and inclusivity in contributions.
    Gao et al.~\cite{gao2023gradientcoin} advocate for a peer-to-peer decentralized network of LLMs, positing that such a structure could enhance robustness and trustworthiness, driving the model towards greater impartiality and openness. Their proposal underscores the potential of decentralized LLMs to overcome the inherent limitations of traditional, centralized models. Similarly, Tang et al.~\cite{tang2023fusionai} contribute to this discourse by highlighting the advantages of distributing certain computational tasks to the client side. This approach not only serves to preserve privacy but also optimizes the processes of pre-training, inference, and fine-tuning of LLMs, offering a novel perspective on the deployment and utilization of LLMs in a manner that prioritizes data security and user privacy.

     \subsection{Security and trust in distributed LLMs}
     Blockchain technology emerges as a pivotal infrastructure in fostering collaborative AI model development and interconnecting LLMs to establish a decentralized AI marketplace, especially the formation of blockchain-based LLM multi-agent systems~\cite{han2024llm}. It primarily offers a mechanism to cultivate trust through the integrity and availability of secure, decentralized knowledge bases during LLM interactions. Moreover, blockchain's contribution to sustainability through incentive mechanisms and reputation systems sets a precedent for further integration and development. Gong~\cite{gong2023dynamic} introduces an innovative concept of a decentralized LLM framework built upon blockchain technology, aiming to imbue LLMs with dynamic capabilities. This model suggests that blockchain not only transitions LLMs from centralized to decentralized architectures but also facilitates their real-time, continuous training. Furthermore, Gong highlights how blockchain can underpin decentralized datasets and economic incentives, thereby fostering an open, collaborative environment for LLM model contributors and validators. This openness and trust paradigm shift, enabled by blockchain, is pivotal for the evolution of dynamic LLMs. Another noteworthy contribution by~\cite{luo2023bc4llm} underscores blockchain's potential in affirming data rights and reshaping profit distribution mechanisms, further emphasizing the technology's transformative impact on LLM ecosystems.
    In the evolving landscape of distributed LLMs, where individual models can communicate and collaboratively enhance their knowledge base, as outlined in the works of~\cite{kuang2023federatedscopellm,zhang2024building}, establishing a secure communication framework becomes paramount. This necessity stems from the potential vulnerabilities where (1) malicious servers might exploit the model exchange to steal parameters or infer sensitive data, and (2) malicious clients could potentially extract information from model parameters or intermediate embeddings.
    Addressing these security concerns within distributed LLM systems poses a significant and open challenge in the field. The question of how to construct a distributed LLM architecture that effectively mitigates these security threats remains largely unresolved. In response to this challenge,~\cite{huang2024fast} has introduced an innovative approach that integrates Trusted Execution Environments (TEEs) to create a secure, distributed LLM framework based on model slicing. Their methodology not only emphasizes the importance of maintaining communication performance and model accuracy but also introduces a novel way of securing the most vulnerable segments of the training model. By deploying the sensitive parts of the model within TEEs at either the sending or receiving ends of the model exchange process, and safeguarding this exchange through robust encryption and decryption mechanisms, Huang et al. offer a promising solution to the dual challenges of ensuring security and preserving the integrity of distributed LLM systems. 


    

    \subsection{Autonomous defense framework}

\begin{figure}[h]
\centering
\includegraphics[width=1\textwidth]{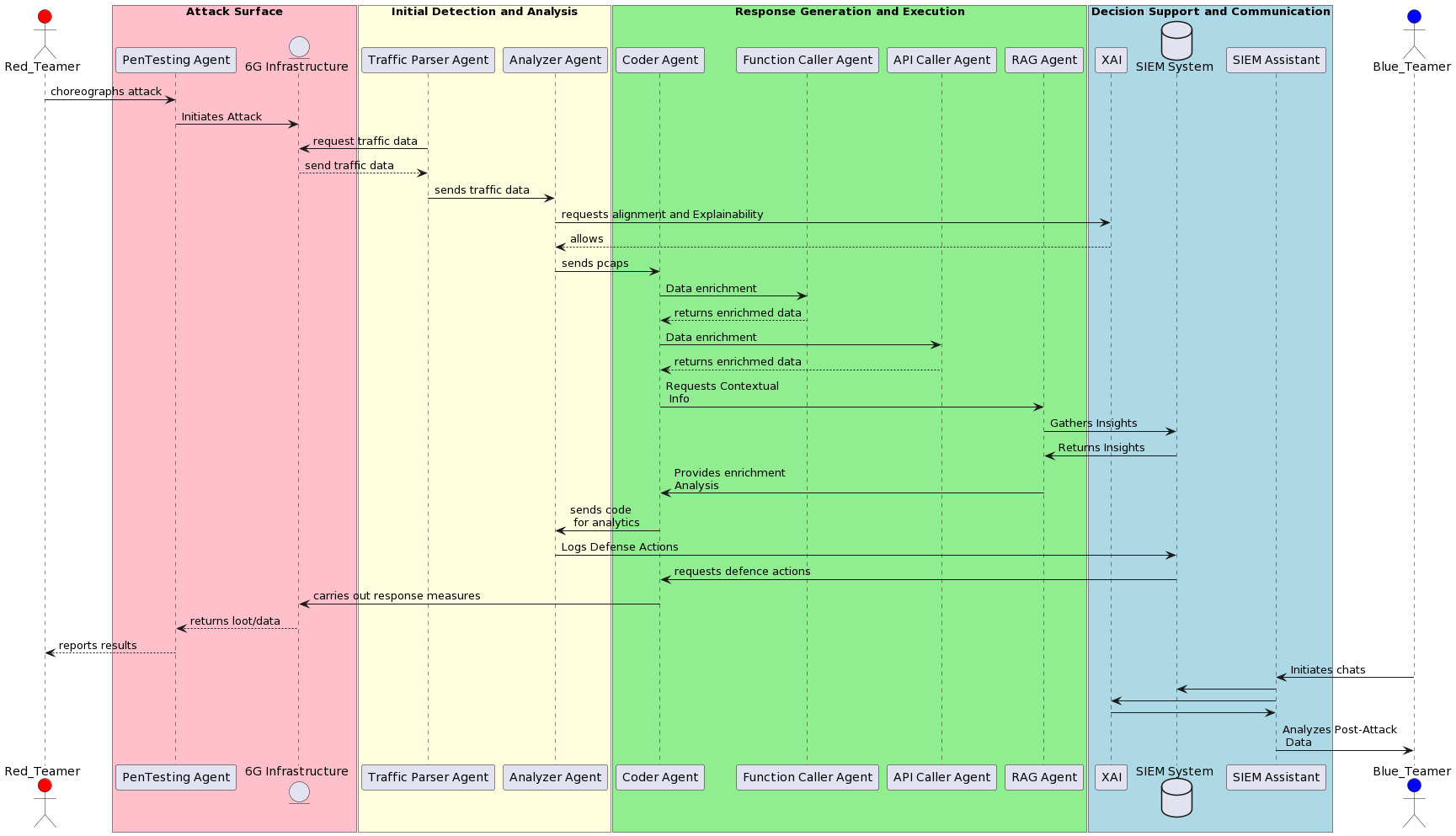}
\caption{Autonomous defense with LLM agent swarms.}
\label{fig:llmswarms}
\end{figure}

    Upon establishing secure and trustworthy LLM and their integration into autonomous swarms of LLM agents, the concept of an autonomous defense framework leveraging LLMs for enhanced cybersecurity emerges as a compelling area of interest. In light of this perspective, we have delineated a comprehensive overview of an autonomous defense framework, employing these swarms of LLM agents, organized into a four-tier taxonomy designed for cyber operations. While this initial framework sketch does not explicitly detail the incorporation of technologies such as blockchain or TEEs, their critical roles in fortifying the framework’s trustworthiness and security are implicitly recognized. The integration of such technologies is envisaged to significantly contribute to the robustness and efficacy of the autonomous defense framework, ensuring a secure and resilient cyber operational environment.
    The sequence diagram depicted in Figure~\ref{fig:llmswarms} represents an illustration of an autonomous defense framework utilizing LLM agent swarms. This framework is designed to identify and respond to adversarial actions instigated by LLM, such as PentestGPT \cite{deng2023pentestgpt} or PAC-GPT \cite{kholgh2023pac}, within cyber environments. Its primary focus is the protection of 6G edge-cloud infrastructure through the deployment of explainable and actionable AI, along with conversational agents designed for human interaction (e.g, HuntGPT \cite{ali2023huntgpt} and Cyber Sentinel \cite{kaheh2023cyber}). The processes are segmented into four main components reflecting the cyber defense lifecycle: Attack Surface, Initial Detection and Analysis, Response Generation and Execution, and Decision Support and Communication.

    \begin{enumerate}
        \item Attack surface: This includes a PenTesting Agent, symbolizing an automated penetration testing entity that interacts with the 6G Infrastructure to simulate attacks or test vulnerabilities. The Red\_Teamer actor initiates the attack, choreographing the PenTesting Agent to initiate attacks on the infrastructure. Choreography in this context can vary from instruction tuning a single local LLM that powers autonomous PenTesting tasks to orchestrating a swarm of red teaming LLMs for more advanced applications.
        
        \item Initial detection and analysis: The Traffic Parser Agent and Analyzer Agent are responsible for the initial detection of potential threats by parsing and analyzing raw traffic data extracted from the infrastructure. The Analyzer Agent may consist of a singular LLM with coding capabilities that have been fine-tuned for a particular task, such as anomaly detection \cite{sheikhi2022novel}, or a collective of agents, each specialized in distinct tasks. This configuration facilitates the early identification of malicious activities and optimizes resource utilization. The Analyzer Agents operate in coordination with the XAI agent, which is tasked with ensuring explainability, interpretability, and addressing general alignment issues. This collaboration ensures that the functionality of the Analyzer aligns with the established operational principles.
        
        \item Response generation and execution: This phase involves several agents including the Coder Agent, Function Caller Agent, API Caller Agent, and Retrieval-Augmented Generation (RAG) Agent. These agents collaborate to generate and execute responses to detected threats by enriching data, gathering contextual information, and performing defensive actions.
        
        \item Decision support and communication: The XAI, SIEM System, and SIEM Assistant form the decision support system. They provide the necessary actionable insights and explanations for the actions taken, ensuring that defense mechanisms are transparent and understandable. They also facilitate communication between the system and the Blue\_Teamer, who is responsible for overseeing the defense operations and analyzing post-attack data.
    \end{enumerate}

\section{Open Research Issues}


Drawing from our comprehensive summarization and analysis, this section concludes with a curated collection of research questions aimed at exploring the secure and safe utilization of LLMs within the 6G edge-cloud continuum. These pivotal research inquiries are categorized into three primary areas: (RQ 1.) Ensuring safety during the training of LLMs; (RQ 2.) Optimizing the integration of LLMs within SecOps to evolve into an effective LLMSecOps framework; and (RQ 3.) Investigating the trustworthiness and security mechanisms in autonomous LLM agent swarms, alongside their potential applications. This structured inquiry seeks to address the critical aspects of LLM deployment, emphasizing safety, efficiency, and security in a rapidly advancing technological landscape.

\textbf{RQ 1:} \textit{What are effective strategies and solutions for enhancing the security and safety of LLM training processes and deployments to mitigate potential vulnerabilities?}\\
The advancement of LLMs hinges on a critical training phase, during which these models ingest vast amounts of data to acquire the knowledge necessary for responding to queries or requests. Consequently, the security vulnerabilities inherent in LLMs often stem from conventional training challenges, including the risk of leaking sensitive information or compromising the integrity of the training models themselves. Furthermore, the deployment phase of LLMs represents an additional dimension of security concern. It necessitates a meticulously crafted design strategy to preempt and mitigate potential vulnerabilities, ensuring the robustness and reliability of the LLM system throughout its lifecycle.




\textbf{RQ 2.1:} \textit{ How can LLMs be optimally utilized within cyber operations to analyze and defend against cyber threats, and how can they be integrated into the SecOps framework to evolve into a comprehensive LLMSecOps approach?}\\
The application of LLMs within the realm of cybersecurity, particularly in cyber operations, holds significant promise, as evidenced by recent studies showcasing LLMs' capabilities in detecting abnormal events~\cite{deng2023pentestgpt,alves2022leveraging,sultana2023towards,guo2021logbert,kaheh2023cyber}. Despite these advancements, there appears to be a gap in integrating these models into a cohesive operational framework. Consequently, the development of a holistic LLMSecOps approach remains an unresolved challenge, inviting further exploration and innovation to bridge this strategic integration gap and fully harness the potential of LLMs in cybersecurity domains.

\textbf{RQ 2.2:} \textit{What are the best practices for efficiently implementing LLMSecOps within the 6G edge-cloud continuum, specifically focusing on IBN, NWDAF, and zero-touch network 6G security features?}\\
Building on the inquiries posited by the preceding research questions, this investigation primarily delves into the application and potential of LLMSecOps within the context of the 6G edge-cloud continuum. In recent times, considerable scholarly effort has been directed toward the deployment and advancement of edge intelligence, with a particular emphasis on securing these developments. However, with the emergent trend of LLMs, the 6G edge-cloud continuum stands on the brink of integrating cutting-edge methodologies inspired by the rapidly evolving domain of LLMSecOps. This promising integration is poised to revolutionize the automation and security configurations within edge-cloud continuum services. Specifically, it heralds significant enhancements in IBN, NWDAF, and zero-touch network security features tailored for the 6G landscape, setting a new benchmark for secure, intelligent network management and operations.






\textbf{RQ 3.1:} \textit{What strategies can be employed to bolster secure connectivity among LLMs to facilitate the formation of effective autonomous LLM agent swarms?}\\
While the idea of leveraging autonomous swarms of LLM agents within a cybersecurity defense framework is compelling, crafting a detailed blueprint for its optimal implementation remains an unresolved challenge. This area of inquiry demands the development of a sophisticated design or strategy that effectively orchestrates LLM agents, ensuring their seamless integration with optimal resources and performance metrics. Such a strategy would not only maximize the efficiency and effectiveness of the autonomous defense framework but also redefine the paradigms of cybersecurity measures. Furthermore, given the nondeterministic nature of LLMs, a key challenge lies in achieving robust coordination and lifecycle management of AI agents to ensure consistent and predictable behavior over time, aligning with the intended functionality.

\textbf{RQ 3.2:} \textit{Can blockchain technology enhance the trustworthiness and security of autonomous LLM agent swarms?}\\
The establishment of trust among interconnected LLM agents emerges as a pivotal concern. Blockchain technology, renowned for its capacity to foster trust through its immutable and transparent nature, presents itself as a prime candidate for creating a trusted network of LLM agents. However, the conceptualization and implementation of a blockchain-powered framework for distributed or decentralized LLMs pose significant challenges that remain largely unaddressed. These challenges are predominantly rooted in the inherent security and performance limitations associated with blockchain technology, which have historically hindered its broader application. Despite these obstacles, the undeniable potential of blockchain technology in a variety of domains underscores a compelling need to explore and develop innovative solutions for integrating blockchain with LLM agent swarms. This exploration is crucial for unlocking new possibilities in creating secure, efficient, and trustworthy digital ecosystems.

\textbf{RQ 3.3:} \textit{Is it feasible for TEE technology to contribute to the security in the formation of autonomous LLM agent swarms?}\\
Within the cybersecurity community, a vigorous debate persists between proponents of traditional cryptographic methods and those advocating for hardware-based security schemes. This debate centers on finding the optimal balance between achieving robust security and maintaining system performance. Hardware-based schemes have shown promising advantages in terms of performance compared to their traditional cryptographic counterparts, particularly in the realms of privacy and security protection. Consequently, the idea of employing hardware-based security mechanisms, such as TEEs, for ensuring the secure formation of LLM agent swarms presents an intriguing prospect. This approach holds considerable potential for enhancing the security posture of LLM agent networks. Despite the apparent promise of this concept, the intricate design and development of TEEs tailored for LLM agent swarms represent a critical next step that remains to be explored, indicating a fertile ground for future research and innovation in secure system architecture.

\textbf{RQ 3.4:} \textit{What constitutes an optimal design for an autonomous defense framework leveraging autonomous LLM agent swarms?}\\
While we have outlined a preliminary concept for an autonomous defense framework utilizing swarms of autonomous LLM agents, achieving an optimal and thoroughly comprehensive design for this framework demands a more detailed exploration. This includes rigorous analyses using formal methodologies to evaluate the framework's integrity, assessments of overhead costs associated with implementing and maintaining the system, and careful consideration of the various potential alternatives for each component within the agent swarms. Such a holistic approach is essential to ensure the framework's effectiveness, efficiency, and adaptability in real-world cybersecurity applications.

\textbf{RQ 3.5:} \textit{How can the deployment of an autonomous defense framework be optimized within the 6G edge-cloud continuum?}\\
Through comprehensive analyses of criteria derived from the autonomous defense framework, its application in bolstering security for next-generation systems—particularly within the context of the 6G era—emerges as a critical next step. This step is underscored by the pivotal role of the edge-cloud continuum, which serves as the cornerstone for systems operating in the 6G era. Specifically, this research question of inquiry aims to identify and develop solutions addressing the challenge of securely deploying an autonomous defense framework atop the foundational edge-cloud continuum in the 6G era. The focus is on achieving an optimal deployment that navigates the complex balance between stringent resource constraints and the demanding security requirements of next-generation telecommunications infrastructure.

\bibliographystyle{plain}
\bibliography{main}

\end{document}